\documentclass[conference]{IEEEtran}
\IEEEoverridecommandlockouts
\usepackage{cite}
\usepackage{amsmath,amssymb,amsfonts}
\usepackage{algorithmic}
\usepackage{graphicx}
\usepackage{textcomp}
\usepackage{xcolor}
\def\BibTeX{{\rm B\kern-.05em{\sc i\kern-.025em b}\kern-.08em
    T\kern-.1667em\lower.7ex\hbox{E}\kern-.125emX}}
\begin{document}

%\title{Conference Paper Title*\\
%{\footnotesize \textsuperscript{*}Note: Sub-titles are not captured in Xplore %and
%should not be used}
%\thanks{Identify applicable funding agency here. If none, delete this.}
%}

\title{Federated Learning for Intrusion Detection in IoT Security: A Hybrid Ensemble Approach}

\author{\IEEEauthorblockN{Sayan Chatterjee and Manjesh K. Hanawal}
\IEEEauthorblockA{
\textit{IEOR, IIT Bombay, Mumbai, India}\\
chatterjee.sayan2014@gmail.com, mhanawal@iitb.ac.in}
}

\maketitle
\begin{abstract}
Critical role of Internet of Things (IoT) in various domains like smart city, healthcare, supply chain and transportation has made them the target of malicious attacks. Past works in this area focused on centralized Intrusion Detection System (IDS), assuming the existence of a central entity to perform data analysis and identify threats. However, such IDS may not always be feasible, mainly due to spread of data across multiple sources and gathering at central node can be costly. Also, the earlier works primarily focused on improving True Positive Rate (TPR) and ignored the False Positive Rate (FPR), which is also essential to avoid unnecessary downtime of the systems. In this paper, we first present an architecture for IDS based on hybrid ensemble model, named PHEC, which gives improved performance compared to state-of-the-art architectures. We then adapt this model to a federated learning framework that performs local training and aggregates only the model parameters. Next, we propose  Noise-Tolerant PHEC in centralized and federated settings to address the label-noise problem. The proposed idea uses classifiers using weighted convex surrogate loss functions. Natural robustness of KNN classifier towards noisy data is also used in the proposed architecture.
Experimental results on four benchmark datasets drawn from various security attacks show that our model achieves high TPR while keeping FPR low on noisy and clean data. Further, they also demonstrate that the hybrid ensemble models achieve performance in federated settings close to that of the centralized settings.
\end{abstract}

\section{Introduction}
\noindent
Internet of Things (IoT) are one of the most effective combinations of physical objects and networks. It involves the uses of embedded systems, wireless networks, machine learning, automation and many other fields. As IoT are becoming critical infrastructure, they can be targets of attackers and need to be proactively protected. Intrusion detection system (IDS) is one of the most common security mechanisms used to thwart malicious attacks. IDS may be anomaly-based or signature-based. Signature-based models generally work well for previously known attacks but not for new/unknown attacks. Machine learning models are increasingly applied in IDS, which work better in detecting previously unknown attacks.\par

Most of the attacks in IoT networks can be broadly categorized into four groups: Denial of Service (DoS), Probe, Remote to Local (R2L), User to Root (U2R). According to the literature \cite{TDTC_2019,TwoTier_2015}, U2R and R2L mimic normal traffic characteristics to a huge extent and hence, difficult to detect. In modern days, a huge number of IoT devices can be connected to a vast network and not all part of the networks see these attacks in equal proportions. For example, some portion of the network may see DoS attack more while the other portion may see more U2R and R2L attacks. Hence data pertaining to these attacks are often collected at a centralized facility to train the models which can detect all type of attacks. However, transferring data to the central node may incur high transfer costs, significant communication overhead, operational constraints \cite{Rhman_20}. Moreover, due to privacy concerns, it may not be practically feasible to collect data from many devices and store them in a single entity. 

%If we consider a single IDS, it may not perform well in maintaining security of huge IoT network, the reason is that it needs huge data and hence a lot of processing time. Also, Hence data pertaining to these attacks are often collected at a centralized facility to train the models which can detect type of attacks. However,in many application, collecting the data at a central place may not be possible due to operational constraints or privacy reasons \cite{}, but can be locally accessed. 
In many IoT networks, the nodes are often equipped with enough computational capabilities such that they can process the data locally. Hence instead of collecting the data at a central node, each source node can train models locally and only share the parameters of the trained models with the central node that can combine the shared parameters appropriately to get the final aggregated model. This federated learning has emerged as a popular method where nodes cooperate, connect and form an aggregated model. This work proposes a Probabilistic Hybrid Ensemble Classification (PHEC) model for detecting threats in IoT networks (in centralized settings) and adapting it to federated settings. The term hybrid in PHEC signifies that multiple classifiers of different types are combined to predict labels. The term probabilistic  indicates that the confidence value in the predicted label is taken into consideration rather than the actual labels of the individual classifiers. Also, Ensemble suggests that a group of classifiers work to give better results. 

\noindent
The main advantage of the PHEC model is its versatility and flexibility -- we can achieve from low False Positive Rate (FPR) to high True Positive Rate (TPR) by tuning the threshold $\gamma$. As is well know in the literature, simultaneously achieving high TPR and low FPR may not be practically feasible -- decreasing FPR reduces TPR and increasing TPR raises FPR\cite{Evil_Tradeoff,Tradeoff_2}. Depending on the type of application, high TPR may be more desirable while FPR up to a threshold is acceptable (fire alarm systems, health and medical diagnosis etc.); in other cases, low FPR is more critical while TPR above a threshold is desirable (justice systems). The proposed model can be adapted to application needs by simply tuning a single hyper-parameter $\gamma$ to achieve the desired trade-off between TPR and FPR.

In the centralized setup of PHEC, we first process the data and perform dimensionality reduction. Two classifiers based on KNN and Random forest algorithms are trained separately using the extracted set of features. Then, the classifiers' outputs are averaged and depending on whether the average is above pre-defined threshold, the final label is decided. %The threshold is tuned using model validation and an optimization criteria that maximizes TPR while keeping FPR within a given tolerance. Most of the past work in IoT security is based on the assumption that the entire data can be stored and processed in a single system. 

PHEC model is also adapted to federated settings with suitable modifications. In federated settings, model parameters (the weights and biases) need to be shared across devices. However, as there are no weights associated with KNN and Random Forest, we use Multi-layer Perceptron (MLP) for training at each node. The final output is the ensembled result of all the Multi-layer Perceptrons.  

In the federated setup, model parameters obtained by averaging the node parameters may not be good in detecting rare attacks. This is because during training, nodes generally face more samples of common attacks like DoS and Probe, which contribute more in defining the aggregated model. So, relatively infrequent attacks become difficult to detect. Another problem that may arise is that different nodes often face different type of attacks (different attacks generally have different characteristics). As an example, let's say node-$1$ faces more DoS attacks and node-$2$ faces mostly Probe attacks. Let $W_{1},W_{2}$ denote model parameters for node-$1$ and node-$2$ respectively. Even though $W_1$ and $W_2$ may be very good individually for DoS and Probe attacks respectively, but the federated model obtained by averaging these parameters $(W_{1}+W_{2})/2$ may not be much effective on either of the attacks! The main reason is the non-iid nature of the training data across different nodes. In our proposed model, we aim to overcome these issues. The  proposed federated PHEC is designed in a way so that each node gets specialized in detecting a particular kind of an attack and then instead of averaging out, we stack them. This kind of aggregation prevents high influence of majority samples on the aggregated model.

Often accuracy of the federated frameworks \cite{Rhman_20} is significantly lower compared to the centralized settings. However, the proposed federated PHEC architecture shows good accuracy and TPR while keeping FPR within tolerable limit.

We next consider the case where labels may be corrupted by noise. Performance of most of the supervised machine learning algorithms drop in the presence of noise. We need Noise Robust Models  to perform well in the presence of label-noise in the training data. It is seen that classifiers that use weighted convex surrogate loss functions, like Biased SVM and Weighted Logistic Regression are provably noise-tolerant \cite{Natarajan_0}. Using weighted version of the convex surrogate loss functions, we modify the PHEC framework and propose Noise-Tolerant PHEC in federated and centralized settings.

%Even though PHEC and the proposed model in federated framework are suited to be applied in different context but they are closely related from an optimization point of view. Here both the models are essentially solving a constrained optimization problem in order to find the threshold confidence value. We will see how these models serve the purpose of a security system by solving this optimization problem. 

In summary, our contributions are as follows:
\begin{itemize}
    \item We develop PHEC architecture for centralized IDS that aims to achieve good TPR while keeping FPR below a user-given permissible limit. The hyper-parameter $\gamma$ allows the system to balance between TPR and FPR. 
    \item To address the privacy concerns related to the data, we adapt PHEC in federated settings. We propose a different kind of model aggregation `FedStacking' here. It is seen from the experiments that federated system works faster than the centralized system.
    \item We evaluate the performance of PHEC on four standard datasets, namely `NSL-KDD', `DS2OS Traffic Traces', `Gas Pipeline Dataset', `Water Tank Dataset'. On the `NSL-KDD' dataset, improvement is up to 10\% TPR in centralized settings and up to 6\% accuracy in Federated settings. We achieved 1\% improvement in TPR on the `DS2OS Traffic Traces' dataset (centralized and federated Settings). We also observed more than 97\% TPR for a minimal amount of FPR on both the `Gas Pipeline' and `Water Tank' datasets in centralized and federated Settings. 
    \item To account for the label-noise problem, we propose Noise-Tolerant PHEC in both centralized and federated settings. We are the first to propose Label-Noise Robust Intrusion Detection System in IoT Security to the best of our knowledge.
\end{itemize}

\subsection{Related Work}
\noindent
Intrusion detection System in Network Security has been well studied in last few years. Existing IDS involves statistical models, machine learning approaches and signal processing models. Classical statistical models like Hidden Markov Model, Bayes Theory\cite{NAIVE_BAYES_2012} have been implemented in this regard. 

\noindent
{\it Centralized IDS setup:} In the domain of Anomaly-based intrusion detection, supervised learning classification models like Random Forest, Naive-Bayes Classifier, SVM, Logistic Regression are often used \cite{detailedKDD2019,Zhang_RF_2008}. Unsupervised approaches\cite{Casas_unsupervised_2012} come handy especially when there is a  shortage of labeled data. A hybrid model based on SVM and hierarchical clustering BIRCH  is proposed in \cite{SVM_BIRCH_HORNG}, which performed well on DoS and Probe attacks but poorly on U2R and R2L attacks. \cite{TDTC_2019} proposed a hybrid model in centralized settings to detect threats in IoT networks. To make the models computationally efficient, \cite{Ambusaidi_2016} applied mutual information to select the optimal number of features for Intrusion Detection System. Most of the earlier works \cite{Xuren_2006,Grouping_NSL_KDD, osana_DDOS_2016} did not include FPR into account and primarily focused on TPR and accuracy. PHEC is different from the earlier works as we focus on optimizing both FPR and TPR simultaneously to the best possible extent. Its performance is good not only in detecting common attacks like DoS but also in detecting rare category of threats like U2R and R2L.
 
\noindent
{\it Federated IDS setup:}  With the advent of fog computing, edge \& cloud computing, distributed frameworks have renewed interest.  Many researchers have explored federated Learning and its applications in recent past\cite{Fed_basics_concept_and_applications} and broader problems and open challenges are discussed in \cite{Challenges_Fed_Learning}. \cite{Rhman_20} proposed IDS in federated settings in IoT networks. They compared on-device training and federated learning framework in their work and carried experimentation on the NSL-KDD dataset\cite{NSL_DATA}. Similar work was also done in \cite{Federated_Tuhin}, where Logistic Regression was used to detect threats in IoT networks. Most of the previous works related to federated learning\cite{Federated_Tuhin,Rhman_20} use a simple weighted average (FedAvg) to update the central server model. Here, we propose a different model aggregation to adapt PHEC in federated settings.

\noindent 
{\it Noise-Tolerant IDS:}  Label-Noise problem is often studied in the standard classification task. The use of weighted convex surrogate loss functions to reduce the effect of label noise is proposed in \cite{Natarajan,Natarajan_0}. \cite{Aritra_Ghosh_2015} provided a sufficient criteria for a loss function to be noise-robust. In this regard, they showed that sigmoid loss, ramp loss, etc. are noise tolerant. \cite{Naresh_2013} showed that risk minimization under 0–1 loss function has impressive noise-tolerance properties. Class-Conditional Noise is also studied by some researchers \cite{CCN_2019}. Here, we are considering the Symmetric Label Noise (SLN) problem only and have proposed Noise Tolerant IDS in centralized and federated settings.\\
\noindent
{\bf Organization of the paper:} Section \ref{sec:ProblemSetup} describes the architecture of the Probabilistic Hybrid Ensemble Classifier. Section \ref{sec:Algo} introduces the proposed algorithm in centralized settings and gives its analysis. Section \ref{sec:Federated Setup} discusses the proposed IDS in federated settings and Section \ref{sec:Noise_PHEC} explains how the PHEC can be adapted to address the Symmetric Label Noise (SLN) problem. Section \ref{sec:Description of the datasets} gives a brief description of the benchmark datasets used for the experiments. Details of the experiments and results are discussed in Section \ref{sec:Implementation and Experimental Results}. Section \ref{sec:Implementation on Noisy Data} discusses the experiments on noisy labeled data. Conclusions with future work is indicated in Section \ref{sec:Conclusion}.

\section{Probabilistic Hybrid Ensemble Architecture}
\label{sec:ProblemSetup}
\noindent
In this section we discuss Probabilistic Hybrid Ensemble Classifier (PHEC) for IDS. PHEC involves an ensemble of classifiers whose predictions are combined to label a given sample as threat or normal. Each classifier is separately trained and are applied on each input parallely.  A similar setup was proposed by Pajouh \cite{TDTC_2019}, where multiple classifiers were used serially and in a sequential manner. The parallel architecture gives the advantage of controlling the False Positive Rate (FPR). We use the standard definitions of  TPR and FPR. Let TP, TN denote the number of True Positives \& True Negatives respectively and FP, FN denote the number of False Positives \& False Negatives respectively. Then, 

\begin{equation*} 
\mbox{TPR}  = \frac{\mbox{TP}}{\mbox{TP+FN}} \quad\mbox{and}\quad \mbox{FPR}  = \frac{\mbox{FP}}{\mbox{FP+TN}}.
\end{equation*}

% \begin{enumerate}
%         \item TPR is the ratio of True Positives (TP) to the total number of actual positive samples, i.e,.
       
%         \item{False Positive Rate}: It is defined as the ratio of False Positives (FP) to the total number of actual negative (non-malicious) samples. 
% \begin{equation} 
% \mbox{FPR} & = \frac{\mbox{FP}}{\mbox{FP+TN}}
% \end{equation}
% \end{enumerate} 

\begin{figure}[h]
\centering
    \includegraphics[scale=.35]{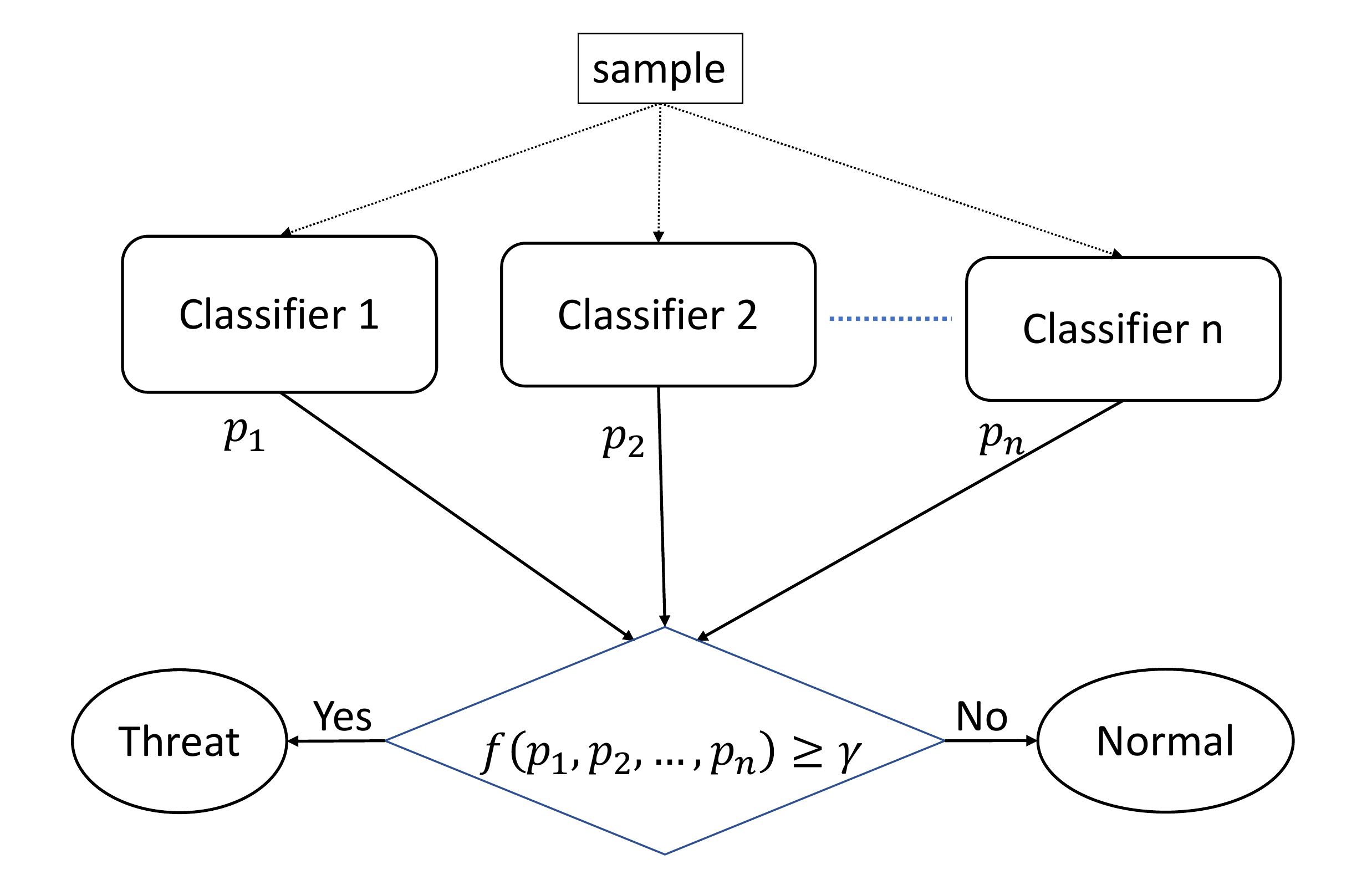}
    \caption{PHEC Architecture}
    \label{fig:training_block_1}
\end{figure}

%PHEC is an ensembled approach that is suited to serve the purpose of a security system and that too in federated as well as in centralized setup. For a test sample, it allows multiple classifiers to predict independently and parallelly the probability that the sample belongs to an attack. A similar setup was proposed by Pajouh \cite{TDTC_2019}, where multiple classifiers were used in series and in a sequential manner. Here the classifiers are implemented parallelly, using the optimization problem that we just referred to. Also, here PHEC is adapted to a federated setup successfully.

PHEC architecture is shown in Figure~\ref{fig:training_block_1}. Let $n$ denote the number of classifiers in PHEC. Each classifier gives a probability or confidence on classifying an input as a threat. For a given sample, let $p_{i}$ for all $i=1,2,\ldots, n$ denote the confidence of the $i$-th classifier in classifying it as a threat. On $\{p_i\}$s, we apply aggregator function $f: [0,1]^n\rightarrow [0,1]$ to get the final score, which is then compared against  a threshold $\gamma$. If $f(p_{1},p_{2},\hdots p_{n}) >= \gamma$, then the sample is labeled as threat, else it is labeled as normal. The parameter $\gamma$ is based on the specified tolerance on FPR.

In security systems, often permissible value of FPR is specified. Let $u$ denote maximum tolerable limit of FPR. Under this circumstances, the objective is to maximize the True Positive Rate (TPR) while keeping the FPR below $u$. For a given $u$, let $g(\gamma)$ denote the TPR and $h(\gamma)$ denote the FPR of PHEC. We set $\gamma \in [0,1]$ such that it maximizes the TPR subjected to the FPR constraint as follows:

% The parameter $\gamma$ is based on  TPR, FPR metrics to measure the performance of our model.
    
% \begin{enumerate}
%         \item {True Positive Rate (TPR)}: It is defined as the ratio of True Positives (TP) to the total number of actual positive samples.
%         It is expressed as,
%         \begin{equation} 
% \mbox{TPR}  = \frac{\mbox{TP}}{\mbox{TP+FN}}
% \end{equation}
%         \item{False Positive Rate}: It is defined as the ratio of False Positives (FP) to the total number of actual negative (non-malicious) samples. 
% \begin{equation} 
% \mbox{FPR} & = \frac{\mbox{FP}}{\mbox{FP+TN}}
% \end{equation}
% \end{enumerate}

% In security systems, often some extent of FPR is tolerable (permissible), say the maximum tolerance limit of FPR is $u$, this value of $u$ is determined by the user. Under this circumstances, security systems need to identify the threats correctly, i.e. maximize TPR. This can be done by suitable tuning of threshold value ($\gamma$). Let's denote FPR as f($\gamma$) and TPR as g($\gamma$), here $\gamma$ is the threshold parameter that is to be found. $\gamma$  is found by solving the following optimization problem.

\begin{align}
\max_{0\leq \gamma \leq 1} \quad & g(\gamma) \nonumber \\
\text{subjected to} \quad  & h(\gamma) \leq u.
\label{eqn:Optimal1}
\end{align}

\section{PHEC model in centralized setup}
\label{sec:Algo}
\begin{figure}[h]
    \centering
    \includegraphics[width=0.35\textwidth]{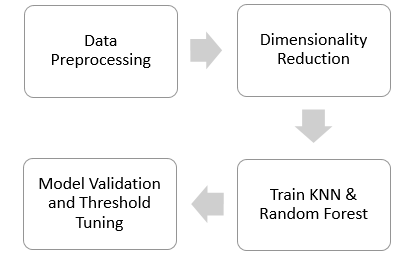}
    \caption{Block Diagram of Training Phase}
    \label{fig:training_block}
\end{figure}

\noindent
This section considers the case where the entire training data can be made available in a single system. We first discuss how each classifier is trained in PHEC and how their outputs are combined to obtain the final label. Figure \ref{fig:training_block} shows the four stages of the training phase in PHEC.

\subsection{Data Prepossessing}

This part involves data cleaning and normalization of features. Data cleaning includes activities like converting categorical variables into numerical values and dropping irrelevant features. Label Encoding is preferred to execute the conversion of categorical variables into suitable numeric form and features are further normalized. It is the process of scaling individual samples to have unit norm. Features, that originally take wide range of values, get scaled into a small range $[-1,1]$ from its original value. The most familiar kind of normalization is `l2' normalization. It creates smoothness and also rotational invariance, which helps further in dimensionality reduction. 
%Due to these advantages of `l2' normalization, it is preferred over `l1' normalization or `max' normalization. 

\subsection{Dimensionality Reduction}
This part uses Principal Component Analysis (PCA) for dimension reduction. This allows the classification framework to use fewer features and helps to make the detection system faster. However, instead of PCA, some other dimension reduction framework like LDA, Autoencoder can be tried out. However, PCA gave better performance on the NSL-KDD dataset as is evident from Table-{\textrm{IX}}. We denote the number of extracted features after dimensionality reduction as $M$. This value of $M$ is a hyperparameter that can be suitably chosen. An useful approach may be to use cross-validation to find the best possible value of $M$. However, this cross-validation itself is quite time-consuming and often, close to the optimal number of extracted dimensions workout well practically.
 
\subsection{Classifier Selection and Training}
In the centralized settings, we restrict to fewer classifiers as each classifier is trained on the entire dataset. We set $n=2$ and use $K$-Nearest Neighbour (KNN) and Random Forest (RF) as the two classifiers. Both the classifiers are trained independently using their probabilistic outputs. One could use other classifiers like SVM, Logistic Regression. However, KNN and Random Forest are preferred due to its simplicity and ensemble nature respectively. Further, the usage of these two classifiers gave better performance than other combinations of classifiers. The aggregator function $f$ is simply the arithmetic mean, i,e, if the prediction probabilities of KNN and Random Forest classifier are $p_{1}$ and $p_{2}$ respectively, then
\begin{center}
$f(p_{1},p_{2})=\frac{(p_{1}+p_{2})}{2}$.
\end{center}
\textbf{Tuning $\gamma$:} Finding optimal $\gamma$ over the range $[0,1]$ is challenging. Instead, we discretize $[0,1]$ into a finite set of values and search for the best value of $\gamma$.  We compute $g(\gamma)$ and $h(\gamma)$ over the validation dataset and obtain optimal value of $\gamma$ by solving optimization problem \ref{eqn:Optimal1} (see Eq.~(\ref{eqn:Optimal1})) over discrete values of $\gamma$. We denote the optimal value of $\gamma$ as $\gamma^*$.
%It is very difficult to find exact optimal solution of $\gamma$ as f($\gamma$) and g($\gamma$) are unknown.

% \begin{align*}
% \max_{0\leq \gamma \leq 1} TPR\\
% \text{s.t.} \quad
% u &\geq FPR \geq 0\\
% 100 &\geq TPR \geq 0\\
% \end{align*}
% Rewriting the optimization problem, let's denote FPR as f($\gamma$) and TPR as g($\gamma$), here $\gamma$ is the threshold parameter that is to be found.
% \begin{align*}
% \max_{0\leq \gamma \leq 1} g(\gamma)\\
% \text{s.t.} \quad
% u &\geq f(\gamma) \geq 0\\
% 100 &\geq g(\gamma) \geq 0\\
% 1 &\geq \gamma \geq 0\\
% \end{align*}

\subsection{Testing Phase}
For a test sample, let the trained KNN and RF predict the chances that it belongs to `Threat' class with confidence $p_1$ and $p_2$, respectively. Then, the test sample is labeled as `Threat' if $f(p_{1},p_{2})\geq \gamma^*$, it is labeled `Normal' otherwise.

\section{PHEC in Federated Settings}
\label{sec:Federated Setup}
\noindent
In this section we discuss the adaptation of PHEC in federated settings. The federated setup in \cite{Federated_Tuhin} allows each node to train on local data that includes samples from different type of attacks and non-malicious samples. Then the node parameters are averaged to form the final aggregated model. The drawback of this model is that the training data are often dominated by samples of common attacks and the parameters of the aggregated central model obtained by averaging the parameters of the nodes do not generalize well to detect rare types of attacks. We overcome this issue by specializing each node for detecting a particular kind of attack and then combining them in a slightly different manner. Our federated setup is based on the idea that each node is trained to detect a particular type of attack and the parameters of each node are used separately at the central node to retain the characteristics of each node.

We group different attacks in training data into $n$ major category based on their characteristics. One such example of grouping of various attacks is shown in Table-\textrm{I}. The value of $n$ depends on the network size, different devices connected to the network etc. However, most of the attacks that IoT networks face generally fall into any one of the $4$ broader groups, namely User to Root (U2R), Remote to Local (R2L), Denial of Service (DoS), and Probe \cite{TDTC_2019}. Hence for practical purposes, $n$ may be chosen as $4$.\\
\begin{table}[!t]
	\caption{Grouping of Different Attacks in IoT Networks} % title of Table
	\centering % used for centering table
	\begin{tabular}{c c} % centered columns (4 columns)
		\hline %inserts double horizontal lines
		Broad Class & Sub-class of attack in training and test data \\ 
		[0.5ex] % inserts table
		%heading
		\hline % inserts single horizontal line
		DoS & back, land, neptune, pod, smurf, teardrop\\
		& apache2, mailbomb, processtable, udpstorm, worm \\ % inserting body of the table
		R2L & ftp write, spy, guess passwd, imap, multhop, phf\\
		& xlock, xsnoop, warezclient, warezmaster, snmpguess\\
		& sendmail, named, snmpgetattack, httptunnel \\
		Probe & ipsweep, nmap, portsweep, satan, mscan, saint\\
		U2R & buffer overflow, perl, loadmodule, rootkit \\
		& ps, sqlattack, xterm\\
		[1ex] % [1ex] adds vertical space
		\hline %inserts single line
	\end{tabular}
	\label{table:attacks grouping} % is used to refer this table in the text
\end{table}

\begin{figure}[h]
	\centering
	\includegraphics[scale=.28]{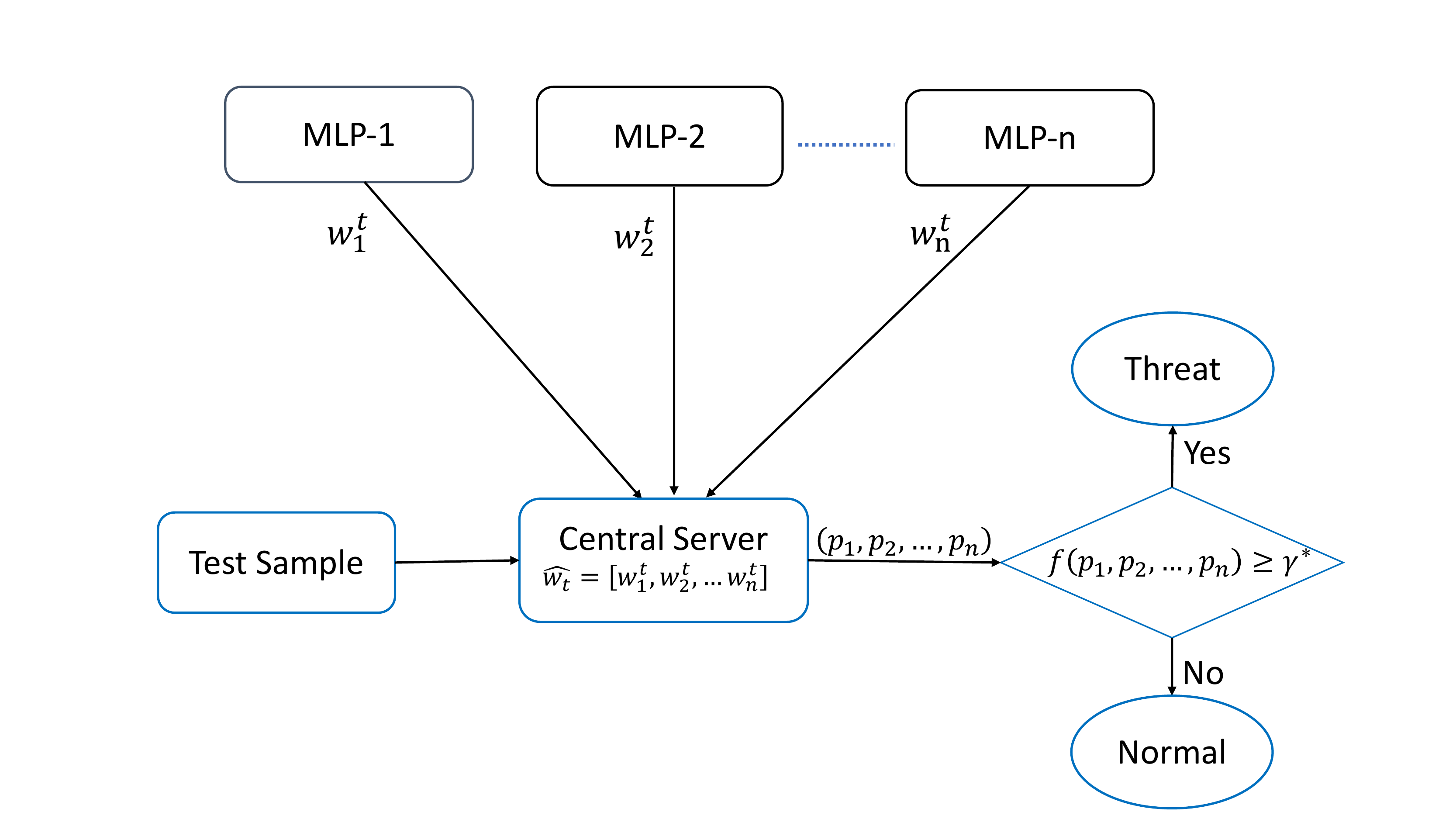}
	\caption{PHEC in federated framework}
	\label{fig:Federated PHEC}
\end{figure}

\noindent
\textbf{Federated Architecture:} The PHEC architecture in the federated setup is shown in Fig. \ref{fig:Federated PHEC}. It consists of $n$ nodes and one central node. Each of the $n$ nodes use a local training dataset to train a Multi-Layer Perceptron (MLP) and share their MLP parameters with the central server where they are aggregated. For each sample data, the aggregated model outputs probability vector $(p_{1},p_{2}\ldots p_{n})$, where $p_i$ corresponds to the probability obtained using the parameters of the $i$ th MLP from $i$ th node. The sample is classified as `Threat' / `Normal'  by thresholding $f(p_{1},p_{2}\ldots p_{n})$ at $\gamma^*$. We next discuss how individual models are trained and model parameters are aggregated.

\noindent
{\bf Dataset Preparation:}
For a completely new system, a test-bed may be setup to collect the data. All the samples belonging to $i$ th type of attack are collected and put in group $D(i)$. Now, the remaining non-malicious samples are added to each group such that the malicious and non-malicious samples are in the ratio $1:1$. Usually, the number of non-malicious samples be much more than the malicious samples and hence multiple datasets with equal proportions of malicious and non-malicious samples can be created.\\
\noindent
\textbf{Classifier Selection:} In the centralized setup of PHEC, Random Forest and KNN were used. However, they cannot be used in the federated setup as they are non-parametric. For better aggregation, the parametric model MLP is a natural choice. Here, each of the $n$ nodes involved trains one MLP each.\\ 
\noindent
\textbf{Label Assignment:} Let $w_i$ denote the set of MLP parameters of node $i$. Each sample passes through each of the $n$ MLPs. Let $p_i$ denote the probability that the sample belongs to malicious class as predicted by the MLP parameters of the $i$ th node. The aggregated model combines the probabilities from $n$ MLPs resulting in a probability vector $(p_{1},p_{2}\ldots p_{n})$. and then apply an aggregate function $f$ as follows:
\begin{center}
    $f(p_{1},p_{2},\cdots p_{n})=max(p_{1},p_{2},\cdots p_{n})$
\end{center}
If $f(p_{1},p_{2},\cdots p_{n})$ exceeds $\gamma^*$, then the sample is labeled as `Threat', else labeled as `Normal', where $\gamma^*$ is a hyperparameter tuned using a validation set.\\
\noindent
Now that the procedure to generate the labels from $\hat{w}_{t}$ and $\gamma^*$ is clear, we will discuss how to find the matrix $\hat{w_{t}}$ and the optimal threshold $\gamma^*$.\\
\noindent
\textbf{Decentralized Training:} Each of the $n$ MLP classifiers is trained using the local dataset available to that particular node ($i$ th node has access to the local dataset $D(i)$). Back-propagation algorithm can be used to tune the parameters (set of weights \& biases) of each of the $n$ MLPs.\\
\noindent
For node $i=1,2, \cdots n$, node parameters are updated in round $t$ as follows:
\begin{enumerate}
\item The gradient is calculated on $D(i)$ as $g_{i}= \nabla F_{i} (w_{i}^{t})$, where $F_{i} (w_{i}^{t})$ is the cross-entropy loss for the $i$-th node with the model parameters $w_{i}^t$.
\item The parameters of $i$-th node at $(t+1)$ iteration are updated as $w_{i}^{t+1} \leftarrow w_{i}^{t}- \eta g_{i}$, where $\eta$ is learning rate.
\end{enumerate}
\textbf{Model Aggregation:} The model aggregation takes place in the central node. Individual nodes can download and use the aggregated model. The most common technique of model aggregation is FedAvg\cite{Federated_Tuhin,Rhman_20}. Instead of FedAvg, we propose and use a stacked aggregation model (FedStacking).
\par In IoT networks, different devices may face different attacks, resulting in non-iid data. Federated architecture with FedAvg Aggregation may not perform well on non-iid data. Secondly, the central model in FedAvg gets more influenced by attacks that frequently occur in training data, resulting in poor detection of rare type of attacks (e.g. U2R,R2L).\\
\noindent
\textbf{FedAvg:} It is the weighted average of all the trained parameters. In the $t$-th iteration, an update of the central server model can be made as,\[\hat{w_{t}}=\sum_{i=1}^{n} \frac{a_i}{a} w_{i}^{t}.\]
Here, $a_i$ denotes the size of the training data for the $i$ th node ($a_{i}=|D(i)|$) and $a=\sum_{i=1}^n a_{i}$.\\
\noindent   
\textbf{FedStacking:} We concatenate the parameters of the trained models (of $n$ nodes) in order to store them efficiently. Consider the matrix $\hat{w}_{t}$ with the number of columns as $n$ such that $i$ th column corresponds to the model parameters $w_{i}^t$. 
\[\hat{w}_{t}= 
\left [w_{1}^t, w_{2}^t, \ldots  w_{n}^t \right ] \]
Now the central node just needs to store the matrix $\hat{w}_{t}$. The matrix $\hat{w}_{t}$ corresponds to the resultant model set of parameters at the end of $t$-th iteration.\\
\noindent  
\textbf{Model Evaluation:} The model is evaluated on a validation dataset. As in the centralized setup, we  discretized the values of $\gamma$ over $[0,1]$ into $S$ uniformly spaced values. Given a predefined value of $u$, the training is performed as follows.\\
For each iteration $t=1,2,\ldots$
\begin{enumerate}
%    \item \textit{Initialization of $t$:} Set $t=1$. 
%    \item \textit{Initialization of $\gamma$:} Set the first item of set $S$ as $\gamma$ 
    \item Using $\hat{w_{t}}$, generate the probability vector $(p_{1},p_{2}\ldots p_{n})$ for each sample in the validation data. 
    \item For each $\gamma \in S$, assign the label to all the samples in the validation data and evaluate associated TPR and FPR.
    \item Find optimal $\gamma$  that maximizes TPR while keeping FPR below $u$ (see Eq.~(\ref{eqn:Optimal1})). Say the optimal value of $\gamma$ in round $t$ is $\gamma^t$.
    \item Repeat untill the stopping criteria is met.
\end{enumerate} 

\noindent  
\textbf{Stopping Criteria:} With the increase in iteration $t$, if it is seen that TPR has saturated/ TPR starts dropping, the training is stopped and the final parameters are stored. As soon as the central node detects the saturation of TPR, the same is communicated to the nodes and local training at all the nodes also stop. If the training is stopped at $t=T$, then $\gamma^T$ is taken as $\gamma^*$ and $\hat{w}_{T}$ gives the final set of parameters. It is  important to carefully choose the stopping point: stopping early may underfit the data whereas stopping late may lead to overfitting of data.

The choice of $n$ plays a major role in the federated settings. There may be trade-off between quicker detection and performance metrics. For large value of $n$, as more number of MLP classifiers get involved, Detection time increases (slower detection). On the other hand, with large $n$, we can make more specific grouping of different attacks, it may lead to better performance metrics.

Some of the benchmark datasets will involve a mix of multiple threats. The usual practice is to divide the benchmark datasets into $n$ groups, followed by pre-processing and supplying it locally to each of the individual node\cite{Federated_Tuhin,Rhman_20}. To evaluate the proposed model, it is customary to split the dataset into multiple datasets and use it as local training data. 

\section{Symmetric Label Noise-Tolerant PHEC}
\label{sec:Noise_PHEC}
\noindent
Label-Noise is a common problem in the domain of supervised Machine-Learning. Symmetric Label Noise (SLN) is defined as the label noise problem where each label is flipped independently with some probability $\rho \in  [0, 0.5)$ and the degree of flipping a label is independent of the actual label of the class. Class-Conditional random noise is the label noise problem where the degree of noise depends on the class in which the instance belongs. Consider a binary classification task with actual label $y\in \{-1,1\}$ and the flipped label $\tilde{y}$, let's define $\rho_{+1}=P(\tilde{y}=-1|y=+1)$, $\rho_{-1}=P(\tilde{y}=+1|y=-1)$. For SLN Noise, $\rho_{+1}=\rho_{-1}=\rho$. For the IDS using supervised methods, it is important to address the issue of the label-noise problem. We modify the proposed PHEC models in centralized and federated setup such that the modified PHEC performs well even in the presence of SLN noise. 

Under the binary classification task, \cite{Natarajan} proposed weighted loss function (convex surrogate functions) in the presence of SLN noise. An unweighted loss function in binary classification (where true label $y \in \{-1,1\}$) takes the form,
\begin{center}
    $l(z,y)=1_{\{ y=1 \}} l_{1}(z) + 1_{\{ y=-1 \}} l_{-1}(z)$
\end{center}
where $z$ represents the margin, $z = yg(x)$,
$g(x)$ corresponds to the function of the specific classifier. $l_{1}(z)$ and $l_{-1}(z)$ indicate the loss function for the true label $y=1$ and $y=-1$ respectively.\\
Proposed $\alpha$-Weighted loss function takes the form,  
\begin{center}
    $l_\alpha(z,y)=(1-\alpha) 1_{\{ y=1 \}} l(z) + \alpha 1_{\{ y=-1\}} l(-z)$
\end{center}
\noindent 
If the loss function given by $l$ is convex, then this $\alpha$-weighted loss function $l_\alpha(z,y)$ (with suitably chosen $\alpha$) turns out to be noise tolerant\cite{Natarajan,Natarajan_0}. As a consequence of this, suitably weighted Biased SVM and weighted Logistic Regression are provably Noise-Tolerant \cite{Natarajan,Natarajan_0}.\\
\noindent
\textbf{Noise-Tolerant PHEC (NT-PHEC) in Centralized Setup}: In the proposed PHEC model in centralized setup, PCA is used for dimension reduction. PCA does not use the label information and hence, label noise-tolerant. For the noise tolerant PHEC, we propose the following modifications to the model proposed in section $\textrm{III}$.

\begin{enumerate}
    \item Random Forest can be replaced by weighted Logistic Regression classifier. Finding  optimal $\alpha$ over the range $[0,1]$ is challenging. Instead, we discretize $[0,1]$ into a finite set of values and search for the best $\alpha$. 
    
    \item The other classifier used in the proposed PHEC is KNN. KNN involving large number of nearest  neighbors i.e. KNN with a higher value of $K$ performs well in the presence of SLN Noise \cite{NOISE_knn}. Intuitively, this is due to the fact that larger the value of $K$, the lesser the effect of label noises on the majority decision that is taken by the KNN classifier. So, we propose to apply the constraint $K\geq 5$ for KNN.\ \
\end{enumerate}
\noindent
\textbf{Noise-Tolerant PHEC (NT-PHEC) in Federated Setup}: For MLP, the resulting cross-entropy loss function may turn out to be non-convex. However, to restore the convexity of the loss function, we propose to limit the number of layers to one input layer and one output layer that consists of a single neuron with a sigmoid activation function. This binary classification unit (with cross-entropy loss) performs similarly to a binary logistic regression classifier and it leads to optimizing over a convex surrogate loss function. Now, $\alpha$-weighted cross-entropy loss can be used as a noise-tolerant convex surrogate loss function. Rest of the architecture remains exactly same as that is proposed in section $\textrm{IV}$.
To find the suitable value of $\alpha$, we discretize $[0,1]$ into a finite set of values and search for the best value of $\alpha$.

The advantages of this proposed model are as follows:
\begin{enumerate}
    \item This proposed architecture is Noise Tolerant. 
    \item For $\alpha=1/2$, weighted cross-entropy loss turns out to be the same as unweighted cross-entropy loss function. With the $\alpha$-weighted loss, $\alpha$ can be tuned in order to find the best possible value of $\alpha$ in the range $[0,1]$. As this range includes $\frac{1}{2}$, range of $\alpha$ in unweighted case is a strict subset of range of $\alpha$ in weighted loss function. So, for the exactly same setup, an MLP classifier, trained with weighted cross-entropy loss (with optimally chosen $\alpha$) is expected to perform at least as good as the MLP classifier (with exactly the same architecture) trained on an unweighted loss function. However, due to discretization of $\alpha$ over the range $[0,1]$, it may be difficult to find out the optimal $\alpha$ practically. Hence, the superiority of $\alpha$-weighted loss functions over unweighted loss functions may not be evident sometimes (if optimal $\alpha$ is not chosen). 
    \item Any class imbalance nature present in the dataset (if exists), can be adjusted by suitable weighting of $\alpha$.
\end{enumerate}
\section{Description of the datasets}
\label{sec:Description of the datasets}
\noindent
In this section, we describe four benchmark datasets that are used to evaluate the  performance of the proposed models.\\
\noindent
{\textbf{NSL-KDD Dataset}}\cite{NSL_DATA,NSL_data_resource}: This NSL-KDD dataset is improved version of the KDD99 dataset. There were some problems pointed out on KDD99, which are later rectified in this NSL-KDD dataset. The modifications in NSL-KDD dataset are as follows:
\begin{enumerate}
    \item Redundant records \& features are removed from the training dataset. Duplicate Records are deleted from the test dataset.
    \item Number of selected records from each difficulty level group is inversely proportional to the percentage of records in KDD99 dataset
    \item The number of samples are quite affordable in both training and test dataset, that makes it relatively easier and cheaper to experiment with.
\end{enumerate}
This dataset is having 3 sections, one is the large `train+' dataset, second one is the `Train+\_20Percent' and last one is the `test+' data. The first two are used for training purposes where `Train+\_20Percent' is a 20\% subset of the `train+' dataset. There are 41 features and one label column in all three datasets. It is important to note that 17 such new type of attacks exist in `test+' dataset that are not present in the training data. This makes NSL-KDD very  relevant in IoT security, as it is very important to detect previously unknown attacks.\par 
\par

\par
\noindent 
Here, this grouping of smaller category attacks into a larger category is essential due to the following reasons:
\begin{enumerate}
    \item For many attacks, very few training samples are available. It is difficult to train a machine learning algorithm on these very few number of samples.
    \item There exist many sub-classes of attacks in the test dataset, which do not exist in training data. So, it is impossible to detect unknown attacks without such grouping.
\end{enumerate}
\noindent
The grouping of sub-classes of attacks into broader categories is given in Table \ref{table:attacks grouping}. Similar grouping was used in \cite{Grouping_NSL_KDD}.\\
\noindent
Even though all these attacks are present in the data, still both training as well as test data are very skewed and the distribution is non-uniform in training and test datasets. There exists class imbalance in NSL-KDD data. More than half of the samples (data points) in the training data belongs to normal traffic and U2R \& R2L are extremely rare in terms of frequency. Even with this low frequency of U2R, R2L attacks, this data is a very realistic and quite accurate representation of distribution of modern-day attacks in IoT networks. This is because the most frequent attack in IoT network is DoS, also U2R and R2L are extremely rare in reality. Even though R2L \& U2R are rare, these occasional attacks may make the entire network vulnerable to malware.\par

\noindent
\textbf{DS2OS Traffic Traces\cite{Dataset_2}:} This is an open source benchmark dataset in IoT security, contributed by \cite{Dataset_2}. The creators captured the data using four simulated IoT sites with different type of services: light controller, thermometer, movement sensors, washing machines, batteries, thermostats, smart doors and smartphones. Each of the sites had a different organization and a different number of services. They created a virtual IoT environment using Distributed Smart Space Orchestration System (DS2OS) for producing synthetic data.

\begin{table}[ht]
\caption{Different Type of Attacks and its Distribution in the Dataset} % title of Table
\centering % used for centering table
\begin{tabular}{c c c c} % centered columns (4 columns)
\hline %inserts double horizontal lines
Attack & Frequency & \% of Data & \% of Anomaly
\\ [0.4ex] % inserts table
%heading
\hline % inserts single horizontal line
Denial of Service & 5780 & 01.61\% & 57.70\%\\
Data Type Probing & 342 & 00.09\% & 03.41\% \\
Malicious Control & 889 & 00.24\% & 08.87\% \\
Malicious Operation & 805 & 00.22\% & 08.03\%\\
Scan & 1547 & 00.43\% & 15.44\%\\
Spying & 532 & 00.14\% & 05.31\%\\
Wrong Setup & 122 & 00.03\% & 01.21\%\\
[1ex] % [1ex] adds vertical space
\hline %inserts single line
\label{table:results_1} % is used to refer this table in the text
\end{tabular}
\end{table}

\begin{table}[ht]
\caption{Features and its Description} % title of Table
\centering % used for centering table
\begin{tabular}{c c c } % centered columns (4 columns)
\hline %inserts double horizontal lines
Feature Number & Features & Type of Data\\ [0.5ex] % inserts table
%heading
\hline % inserts single horizontal line
1 & Source ID & Nominal\\
2 & Source Address & Nominal\\
3 & Source Type & Nominal\\
4 & Source Location & Nominal\\
5 & Destination Service Address & Nominal\\
6 & Destination Service Type & Nominal\\
7 & Destination Location & Nominal\\
8 & Accessed Node Address & Nominal\\
9 & Accessed Node Type & Nominal\\
10 & Operation & Nominal\\
11 & Value & Continuous\\
12 & Timestamp & Discrete\\
13 & Normality & Nominal\\
[1ex] % [1ex] adds vertical space
\hline %inserts single line
\label{table:results_1} % is used to refer this table in the text
\end{tabular}
\end{table}
\noindent
Table-\textrm{II} describes the different types of attacks and their distributions. Clearly, majority of the attack samples belong to the `Denial of Service' type. Table-\textrm{III} describes different features and their corresponding datatypes. Here the target variable is `Normality'.

\noindent
\textbf{Gas Pipeline Data (SCADA Traffic and Payload Datasets)\cite{DATA3_14}:} This is a freely available dataset collected in Gas Pipeline. It was generated from network flow records that was captured with a serial port data logger. Here two categories of features are present, namely network traffic features and payload content features. Network traffic features include the device address, function code, length of packet, packet error checking information and time intervals between packets. While ``network traffic features'' account for the communication patterns in networks, ``Payload content features'' takes care of the system's current state. 

\noindent
First two datasets\cite{NSL_DATA,Dataset_2} include features only related to network traffic. However, this Gas Pipeline Data includes features involving payload patterns as well as Network Traffic. 

\noindent
\textbf{Water Tank Data (SCADA Traffic and Payload Datasets)\cite{DATA3_14}:} This dataset is quite similar to the Gas Pipeline Data, both these datasets include Traffic and Payload Data in a SCADA network. However, this dataset is collected in Water Storage Tank instead of a Gas Pipeline system. 
\section{Implementation \& Experimental Results}
\label{sec:Implementation and Experimental Results}
\noindent
The experiments are conducted in Google Colab using Python programming language, running on a personal computer. All the performance metrics are expressed in percentages. 

\subsection{NSL-KDD Dataset}
\subsubsection{\normalfont{\textbf{PHEC in Centralized Setup}}}
For the experimental results in Table-\ref{table:results_1}, extracted number of dimensions considered are 20. In Table-\ref{table:results_1}, models are trained on smaller training dataset (`Train+\_20 Percent'). It can be observed that PHEC is the best performing model in terms of detecting threats and that by quite significant margin (maximum TPR obtained using PHEC is atleast 10\% more compared to any other model).
\begin{table}[ht]
\caption{Results on `test+' dataset (trained on `train+\_20percent' data)} % title of Table
\centering % used for centering table
\begin{tabular}{c c c c c c} % centered columns (4 columns)
\hline %inserts double horizontal lines
Model & $\gamma^*$ & TPR & FPR & Accuracy & $u$\\ [0.5ex] % inserts table
%heading
\hline % inserts single horizontal line
PHEC & 0.015 & \textbf{94.6} & 12.34 & 91.6 & 13\\
PHEC & 0.030 & 92.28 & 9.94 & 91.34 & 10\\ 
PHEC & 0.060 & 88.04 & 6.22 & 90.77 & 6.5\\
PHEC & 0.065 & 86.98 & 5.39 & 90.27 & 6\\
PHEC & 0.075 & 86.38 & 5.25 & 89.98 & 5.25\\
TDTC\cite{TDTC_2019} & N/A & 84.82 & 5.56 & N/A & N/A \\
Two-Tier\cite{TwoTier_2015}& N/A & 83.24 & \textbf{4.83} & N/A & N/A\\
Naive Bayes\cite{detailedKDD2019} & N/A & 76.56 & N/A & N/A & N/A\\
Random Forest\cite{detailedKDD2019} & N/A & 80.67 & N/A & N/A & N/A\\
SVM\cite{detailedKDD2019} & N/A & 69.52 & N/A & N/A & N/A\\
Decision Trees(J48)\cite{detailedKDD2019} & N/A & 81.05 & N/A & N/A & N/A\\
[1ex] % [1ex] adds vertical space
\hline %inserts single line
\label{table:results_1} % is used to refer this table in the text
\end{tabular}
\end{table}

\noindent
If TDTC\cite{TDTC_2019} and PHEC (at $\gamma^*=0.075$) are compared in Table-\ref{table:results_1}, it can be seen that for PHEC, results have significantly improved in terms of both FPR and TPR. It indicates direct superiority of PHEC over TDTC\cite{TDTC_2019}. Compared to Two-Tier model\cite{TwoTier_2015}, TPR in PHEC (at $\gamma^*=0.075$) improved by 3.14\% at the cost of only additional 0.42\% FPR. Also, it is seen that as $\gamma^*$ drops, desired metric TPR gets improved but that happens at the cost of larger value of FPR.
\begin{table}[ht]
\caption{Results on `test+' dataset (model trained on larger training dataset)} % title of Table
\centering % used for centering table
\begin{tabular}{c c c c c c} % centered columns (4 columns)
\hline %inserts double horizontal lines
Model & $\gamma^*$ & TPR & FPR & Accuracy & $u$\\ [0.5ex] % inserts table
%heading
\hline % inserts single horizontal line
PHEC & 0.010 & \textbf{95.85} & 12.70 & 92.17 & 13\\
PHEC & 0.020 & 93.07 & 10.91 & 91.41 & 11\\
PHEC & 0.045 & 90.875 & 6.8 & 91.89 & 7\\
PHEC & 0.050 & 90.06 & 6.44 & 91.52 & 6.5\\
SVM-IDS\cite{Zhang_RF_2008} & N/A & 82 & 15 & N/A & N/A\\
Two-Tier\cite{TwoTier_2015} & N/A & 81.97 & 5.44 & N/A & N/A\\
TDTC\cite{TDTC_2019} & N/A & 84.86 & \textbf{4.86} & N/A & N/A\\
SOM-IDS\cite{D'Orazio_2016} & N/A & 75.49 & N/A & N/A & N/A\\
[1ex] % [1ex] adds vertical space
\hline %inserts single line
\label{table:results_2} % is used to refer this table in the text
\end{tabular}
\end{table}

\noindent
The results in Table-\ref{table:results_2} are obtained when models are trained on larger training dataset. Number of extracted dimensions considered are 16. If results of PHEC are compared with that of Two-Tier\cite{TwoTier_2015} and TDTC\cite{TDTC_2019} models, it is seen that the TPR has improved by a large margin at the expense of very small increase to the metric FPR. Compared to the Two Tier model\cite{TwoTier_2015}, TPR has improved by 8.09\% at the expense of 1\% extra FPR in PHEC ($\gamma^*=0.050$). Also, compared to the TDTC\cite{TDTC_2019}, TPR has improved by by 5.20\% at the expense of only 1.58\% extra FPR in PHEC ($\gamma^*=0.050$).\\
\noindent
\textbf{Classification Results (For Each Category of Attack):} We evaluate the effectiveness of the proposed PHEC algorithm on common as well as on uncommon category of attacks. The results show that PHEC is the best performing model for common (Probe) as well as uncommon attacks (U2R, R2L).\\

For Table-\textrm{VI}, FPR is  $5.39$\%, $\gamma^*=0.065$ and $u=6\%$. 

\begin{table}[ht]
\caption{results on `test+' data for each category of attack (model trained on `Train+\_20Percent' dataset)} % title of Table
\centering % used for centering table
\begin{tabular}{c c c c c c c} % centered columns (4 columns)
\hline %inserts double horizontal lines
Attack & Total & Detected as Threat & Not Detected & TPR\\ [0.5ex] % inserts table
%heading
\hline % inserts single horizontal line
R2L & 2885 & 1571 & 1314 & 54.45\\ % inserting body of the table
U2R & 67 & 53 & 14 & 79.10\\
DoS & 7460 & 7124 & 336 & 95.50\\
Probe & 2421 & 2415 & 6 & 99.75\\
[1ex] % [1ex] adds vertical space
\hline %inserts single line
\end{tabular}
\end{table}
\label{table:Results_3} % is used to refer this table in the text

\noindent
For Table-\textrm{VII}, FPR is $6.44\%$, $\gamma^*=0.05$ and $u=6.5\%$. 
\begin{table}[ht]
\caption{results on `test+' data for each category of attack (model trained on larger training dataset)} % title of Table
\centering % used for centering table
\begin{tabular}{c c c c c c c} % centered columns (4 columns)
\hline %inserts double horizontal lines
Attack & Total & Correctly Classified & Misclassified & TPR\\ [0.5ex] % inserts table
%heading
\hline % inserts single horizontal line
R2L & 2885 & 1957 & 928 & 67.83\\ % inserting body of the table
U2R & 67 & 40 & 27 & 59.70\\
DoS & 7460 & 7154 & 306 & 95.90\\
Probe & 2421 & 2397 & 24 & 99\\
[1ex] % [1ex] adds vertical space
\hline %inserts single line
\end{tabular}
\end{table}
\label{table:Results_4} % is used to refer this table in the text
\begin{table}[ht]
\caption{Comparative analysis of TPR on `test+' dataset} % title of Table
\centering % used for centering table
\begin{tabular}{c c c c c c} % centered columns (4 columns)
\hline %inserts double horizontal lines
Algorithm & Trained on & Probe & DoS & U2R & R2L\\ 
[0.5ex] % inserts table
%heading
\hline % inserts single horizontal line
PHEC & Train+ & 99 & 95.90 & 59.70 & \textbf{67.83}\\
PHEC & Train+\_20\% & \textbf{99.75} & 95.50 & \textbf{79.10} & 54.45\\
SVM with BIRCH\cite{SVM_BIRCH_HORNG} & N/A & 99.5 & \textbf{97.5} & 28.8 & 19.7\\
TDTC\cite{TDTC_2019} &  N/A & 87.32 & 88.20 & 70.15 & 42\\
Two-tier\cite{TwoTier_2015} & N/A & 79.76 & 84.68 & 67.16 & 34.81\\
Association IDS \cite{detailedKDD2019} & N/A & 96.8 & 74.9 & 0.79 & 0.38\\
HFR-MLR \cite{HFRMLR_2016} & N/A & 80.2 & 89.70 & 29.50 & 34.20\\
ESC-IDS\cite{NEURO_FUZZY_Toosi} & N/A & 99.5 & 84.1 & 31.5 & 14.1\\
[1ex] % [1ex] adds vertical space
\hline %inserts single line
\label{table:Attackwise comparative summary} % is used to refer this table in the text
\end{tabular}
\end{table}

\noindent
\textbf{Effect of Dimension Reduction on Results:}
In the proposed PHEC algorithm, different dimension reduction techniques can be applied. We implement nonlinear dimensionality reduction like Autoencoder and linear dimensionality reduction like Linear Discriminant Analysis (LDA) along with PCA in the proposed PHEC setup. In Table-\textrm{IX}, it can be seen that performance of PHEC with PCA is way better than other dimension reduction techniques. 
\begin{table}[ht]
\caption{Effect of Dimension Reduction on Results} % title of Table
\centering % used for centering table
\begin{tabular}{c c c c c c} % centered columns (4 columns)
\hline %inserts double horizontal lines
Dimension Reduction & $\gamma^*$ & FPR & TPR\\ [0.5ex] % inserts table
%heading
\hline % inserts single horizontal line
Autoencoder & 0.010 & 13.625 & 89.74\\
Autoencoder & 0.020 & 11.46 & 84.92\\
Autoencoder & 0.045 & 9.88 & 78.10\\
Autoencoder & 0.050 & 9.70 & 77.51\\
LDA & 0.010 & 13.29 & 72.25\\
LDA & 0.020 & 13.23 & 72.13\\
LDA & 0.045 & 13.23 & 72.13\\
LDA & 0.050 & 13.23 & 72.13\\
PCA & 0.010 & 12.7 & \textbf{95.85}\\
PCA & 0.020 & 10.91 & 93.07\\
PCA & 0.045 & 6.80 & 90.875\\
PCA & 0.050 & \textbf{6.44} & 90.06\\
PCA+LDA & 0.010 & 13.15 & 72.67\\
PCA+LDA & 0.020 & 13.09 & 72.48 \\
PCA+LDA & 0.045 & 13.09 & 72.48 \\
PCA+LDA & 0.050 & 13.09 & 72.48 \\
[1ex] % [1ex] adds vertical space
\hline %inserts single line
\end{tabular}
\end{table}
\label{table:Results_dim} % is used to refer this table in the text

\subsubsection{\normalfont{\textbf{PHEC in Federated Setup}}}
\noindent
Table-\textrm{X} shows the performance of proposed PHEC Model in federated settings. Even though performance in Federated Settings dropped compared to PHEC in centralized settings, still very high TPR along-with decent accuracy can be obtained here. 
\begin{table}[ht]
\caption{Experimental Results of PHEC on NSL-KDD Dataset (`test+' Dataset) in Federated Settings} % title of Table
\centering % used for centering table
\begin{tabular}{c c c c c c c c} % centered columns (4 columns)
\hline %inserts double horizontal lines 
$n$ & $u$ & $\gamma^*$ & Accuracy & TPR & Precision & FPR & Training Data\\ [0.5ex] % inserts table
%heading
\hline % inserts single horizontal line
4 & 10 & 0.75 & 88.42 & 96.67 & 85.04 & 9.68 & `Train+'\\
4& 9 & 0.875 & 87.27 & 93.16 & 85.72 & 8.83 & `Train+'\\
4 & 10 & 0.015 & 83.78 & 88.46 & 83.92 & 9.64 & `Train+\_20Percent'\\% inserting body of the table
[1ex] % [1ex] adds vertical space
\hline %inserts single line
\end{tabular}
\end{table}
\noindent
In a recent work on IoT security \cite{Rhman_20}, test accuracy reported on NSL-KDD dataset in Federated settings was around 82\% but significantly improved performance is obtained here. 
%\end{enumerate}
%\end{enumerate}
\subsection{\textit{DS2OS Traffic Traces Dataset}}
\begin{enumerate}
    \item {\textbf{PHEC in Centralized Setup:}}
\begin{table}[ht]
\caption{Experimental Results on `DS2OS Traffic Traces' Dataset (Training Data) in Centralized Settings} % title of Table
\centering % used for centering table
\begin{tabular}{c c c c c c c c c} % centered columns (4 columns)
\hline %inserts double horizontal lines
Model & Accuracy & TPR & FPR & F1-Score\\ [0.5ex] % inserts table
%heading
\hline % inserts single horizontal line
PHEC &  \textbf{99.97} & \textbf{100} & 0.026 & \textbf{99}\\ % inserting body of the table
Logistic Regression\cite{bangladesh_19} & 98.3 & 98 & N/A & 98\\ 
SVM\cite{bangladesh_19}	& 98.2 & 98 & N/A & 98\\ 
Decision Tree\cite{bangladesh_19} & 99.4 & 99 & N/A & \textbf{99}\\
Random Forest\cite{bangladesh_19} & 99.4 & 99 & N/A & \textbf{99}\\ 
Neural Net\cite{bangladesh_19} & 99.4 & 99 & N/A & \textbf{99}\\
[1ex] % [1ex] adds vertical space
\hline %inserts single line
\end{tabular}
\end{table}
Table-\textrm{XI} \& \textrm{XII} presents comparative results on training and test dataset of `DS2OS Traffic Traces' respectively. Here, the maximum tolerable limit of FPR $u$ is fixed at $1\%$.

\begin{table}[ht]
\caption{Experimental Results on `DS2OS  Traffic  Traces' Dataset (Test Data) in Centralized Settings} % title of Table
\centering % used for centering table
\begin{tabular}{c c c c c c c c} % centered columns (4 columns)
\hline %inserts double horizontal lines
Model & Accuracy & TPR & FPR & F1-Score\\ [0.5ex] % inserts table
%heading
\hline % inserts single horizontal line
PHEC & \textbf{99.98} & \textbf{100} & 0.016 & \textbf{99.3}\\ % inserting body of the table
Logistic Regression\cite{bangladesh_19} & 98.3 & 98 & N/A & 98\\ 
SVM\cite{bangladesh_19}	& 98.2 & 98 & N/A & 98\\ 
Decision Tree\cite{bangladesh_19} & 99.4 & 99 & N/A & 99\\
Random Forest\cite{bangladesh_19} & 99.4 & 99 & N/A & 99\\ 
Neural Net\cite{bangladesh_19} & 99.4 & 99 & N/A & 99\\
[1ex] % [1ex] adds vertical space
\hline %inserts single line
\end{tabular}
\end{table}
\item{\textbf{PHEC in Federated Setup:}} PHEC achieves more than $98\%$ accuracy on `DS2OS Traffic Traces' data in federated settings. 
\begin{table}[ht]
\caption{Experimental Results on `DS2OS  Traffic  Traces' Dataset (Test Data) in Federated Settings} % title of Table
\centering % used for centering table
\begin{tabular}{c c c c c c c c c} % centered columns (4 columns)
\hline %inserts double horizontal lines
$u$ & $n$ & $\gamma^*$ & Accuracy & TPR & FPR & Precision & F1-score\\ [0.5ex] % inserts table
%heading
\hline % inserts single horizontal line
2 & 7 & 0.925 & 98.27 & 97.69 & 1.674 & 61.5 & 75.49\\
[1ex] % [1ex] adds vertical space
\hline %inserts single line
\end{tabular}
\end{table}
\end{enumerate}

\subsection{Gas pipeline Data (SCADA Traffic and Payload Datasets)}
\begin{enumerate}
    \item {\textbf{PHEC in Centralized Setup:}} Table-\textrm{XIV} presents the experimental results on `Gas Pipeline' dataset. Results indicate the excellence of performance of PHEC in centralized settings.   
\begin{table}[ht]
\caption{Experimental Results on `Gas Pipeline' Data in Centralized Settings (Test Data)} % title of Table
\centering % used for centering table
\begin{tabular}{c c c c c c c c c c} % centered columns (4 columns)
\hline %inserts double horizontal lines
$u$ & $\gamma^*$ & Accuracy & TPR & FPR & Precision & F1-score\\ [0.5ex] % inserts table
%heading
\hline % inserts single horizontal line
2 & 0.3 & 98.38 & 97.56 & 1.15 & 97.98 & 97.77\\
[1ex] % [1ex] adds vertical space
\hline %inserts single line
\end{tabular}
\end{table}
\item{\textbf{PHEC in Federated Setup:}} \textrm{Table-XV} shows that the performance observed in Federated settings (in PHEC) is marginally better than centralized settings! Generally, centralized settings gives better performance metrics, here we observe slight exception in results.  
\begin{table}[ht]
\caption{Experimental Results on `Gas Pipeline' Data (Test Data) in Federated Settings} % title of Table
\centering % used for centering table
\begin{tabular}{c c c c c c c c} % centered columns (4 columns)
\hline %inserts double horizontal lines
$u$ & $n$ & $\gamma^*$ & Accuracy & TPR & FPR & Precision & F1-score\\ [0.5ex] % inserts table
%heading
\hline % inserts single horizontal line
1 & 7 & 0.95 & 98.71 & 97.47 & 0.36 & 98.99 & 98.22\\
[1ex] % [1ex] adds vertical space
\hline %inserts single line
\end{tabular}
\end{table}
\end{enumerate}

\subsection{Water Tank (SCADA Traffic and Payload Data)}
\begin{enumerate}
    \item {\textbf{PHEC in Centralized Setup:}}
Here PHEC is implemented on `Water Tank' dataset in centralized settings and results are presented in Table-\textrm{XVI}. Results show that for a very less FPR ($<1\%$), obtained TPR and Accuracy are very high! (Both are $>99\%$.)
\begin{table}[ht]
\caption{Experimental Results on `Water Tank' Data (Test Data) in Centralized Settings} % title of Table
\centering % used for centering table
\begin{tabular}{c c c c c c c c c} % centered columns (4 columns)
\hline %inserts double horizontal lines
$u$ & $\gamma^*$ & Accuracy & TPR & FPR & Precision & F1-score\\ [0.5ex] % inserts table
%heading
\hline % inserts single horizontal line
1 & 0.25 & 99.11 & 99.37 & 0.99 & 97.40 & 98.37\\
[1ex] % [1ex] adds vertical space
\hline %inserts single line
\end{tabular}
\end{table}

\item{\textbf{PHEC in Federated Setup:}} Table-\textrm{XVII} shows that even though performance of PHEC dropped in federated settings, still the performance is quite good.  
\begin{table}[ht]
\caption{Experimental Results on `Water Tank' Data (Test Data) in Federated Settings} % title of Table
\centering % used for centering table
\begin{tabular}{c c c c c c c c c c} % centered columns (4 columns)
\hline %inserts double horizontal lines
$u$ & $n$ & $\gamma^*$ & Accuracy & TPR & FPR & Precision & F1-score\\ [0.5ex] % inserts table
%heading
\hline % inserts single horizontal line
6 & 7 & 0.95 & 93.74 & 97.41 & 5.56 & 82.34 & 89.24\\
[1ex] % [1ex] adds vertical space
\hline %inserts single line
\end{tabular}
\end{table}
\end{enumerate}

\subsection{Comparison of Time in Centralized and Federated Settings}
\noindent
Here we compare the time required to detect a test sample in centralized and federated settings. It can be seen that time required is (10-100) times less in federated setup. This is very useful in IoT security as often, intrusions have substantial effect on system and detecting at earlier stage makes sure that immediate actions can be taken.
\begin{table}[ht]
\caption{Comparison of Time in Centralized and Federated Settings} % title of Table
\centering % used for centering table
\begin{tabular}{c c c c c c c c c} % centered columns (4 columns)
\hline %inserts double horizontal lines
Dataset & Settings & Time/Instance\\ [0.5ex] % inserts table
%heading
\hline % inserts single horizontal line
NSL-KDD & Centralized &  $9.02*10^{-6}$s\\
NSL-KDD & Federated & $1.14*10^{-8}s$\\
DS2OS & Centralized & $37.1 * 10^{-6}s$\\
DS2OS & Federated & $0.736*10^{-8}s$\\
Gas Pipeline Data & Centralized & $6.96*10^{-6}s$\\
Gas Pipeline Data & Federated & $0.882*10^{-8}s$\\
Water Tank Data & Centralized & $10.6*10^{-6} s$\\
Water Tank Data & Federated & $8.89*10^{-8}s$\\
[1ex] % [1ex] adds vertical space
\hline %inserts single line
\end{tabular}
\end{table}

\section{Implementation and Results on Noisy Data}
\label{sec:Implementation on Noisy Data}
\subsection{Centralized Settings}
\noindent
Degree of noise (in Percentage) is represented by $\rho$ and we consider, $\rho \in \{10,20\}$.\par
\noindent
The following steps are taken to generate noisy training labels. 
\begin{enumerate}
    \item For each instance of the training data, generate one uniform random number $v$ that lies in-between $[0,1]$.
    \item If $100v\leq \rho$, then change the label of that instance. Otherwise, keep it as it is. 
\end{enumerate}  
Now, the training data is having $\rho$\% of the training samples with noisy labels. 

\begin{table}[ht]
\caption{Experimental Results on Noisy Data using Noise Tolerant PHEC in Centralized Settings} % title of Table
\centering % used for centering table
\begin{tabular}{c c c c c} % centered columns (4 columns)
\hline %inserts double horizontal lines
Dataset & $\rho$ & Accuracy & FPR & TPR\\ [0.5ex] % inserts table
%heading
\hline % inserts single horizontal line
NSL-KDD & 10 & 83.22 & 11.05 & 78.89\\
NSL-KDD & 20 & 81.02 & 11.12 & 75.07\\
DS2OS & 10 & 99.96 & 0.0186 & 98.46\\
DS2OS & 20 & 95.13 & 4.87 & 95.64\\
Gas Pipeline & 10 & 97.48 & 1.66 & 95.98\\
Gas Pipeline & 20 & 92.74 & 6.844 & 92.015\\
Water Tank & 10 & 97.83 & 0.73 & 93.87\\
Water Tank & 20 & 93.74 & 6.61 & 94.71\\
[1ex] % [1ex] adds vertical space
\hline %inserts single line
\end{tabular}
\end{table}

\noindent
The performance of Noise Tolerant PHEC (NT-PHEC) on noisy and clean data are tabulated in Table-\textrm{XIX} and Table-\textrm{XX} respectively. 
\begin{table}[ht]
\caption{Experimental Results on Clean Data using Noise Tolerant PHEC in Centralized Settings} % title of Table
\centering % used for centering table
\begin{tabular}{c c c c c} % centered columns (4 columns)
\hline %inserts double horizontal lines
Dataset & Accuracy & FPR & TPR\\ [0.5ex] % inserts table
%heading
\hline % inserts single horizontal line
NSL-KDD & 83.77	& 3.2 & 73.92\\
DS2OS & 99.97 &	.01 & 98.56\\
Gas Pipeline & 98.33 & 0.64	& 96.53\\
Water Tank & 98.64 & 0.66 & 96.72\\
[1ex] % [1ex] adds vertical space
\hline %inserts single line
\end{tabular}
\end{table}

\subsection{Federated Settings}
\noindent
The results obtained using Noise Tolerant PHEC (NT-PHEC) on noisy as well as on clean data are presented in this section.

\begin{table}[ht]
\caption{Experimental Results on Noisy Data using Noise Tolerant PHEC in Federated Settings} % title of Table
\centering % used for centering table
\begin{tabular}{c c c c c} % centered columns (4 columns)
\hline %inserts double horizontal lines
Dataset & $\rho$ & Accuracy & FPR & TPR\\ [0.5ex] % inserts table
%heading
\hline % inserts single horizontal line
NSL-KDD & 10 & 76.48 & 19.12 & 92.30\\
NSL-KDD & 20 & 74.22 & 21.44 & 92.39\\
DS2OS & 10 & 91.66 & 7.34 & 63.54\\
DS2OS & 20 & 90.48 & 8.50 & 62.87\\
Gas Pipeline & 10 & 94.26 & 1.282 & 87.79\\
Gas Pipeline & 20 & 93.988 & 1.28 & 87.046\\
Water Tank & 10 & 90.47 & 0.0017 & 64.21\\
Water Tank & 20 & 90.22 & 0.28 & 64.35\\
[1ex] % [1ex] adds vertical space
\hline %inserts single line
\end{tabular}
\end{table}

\begin{table}[ht]
\caption{Experimental Results on Clean Data using Noise Tolerant PHEC in Federated Settings} % title of Table
\centering % used for centering table
\begin{tabular}{c c c c c} % centered columns (4 columns)
\hline %inserts double horizontal lines
Dataset & Accuracy & FPR & TPR\\ [0.5ex] % inserts table
%heading
\hline % inserts single horizontal line
NSL-KDD & 83.65	& 15.86 & 99.15\\
DS2OS & 94.224 & 5.22 & 79.89\\
Gas Pipeline & 94.75 & 1.28 & 89.13\\
Water Tank & 90.48 & 0.0067 & 64.28\\
[1ex] % [1ex] adds vertical space
\hline %inserts single line
\end{tabular}
\end{table}

\subsection{Comparison of Results on Noisy and Clean Data}
\noindent
Here, we compare the accuracy obtained using NT-PHEC and PHEC in both centralized and federated settings. 
\begin{table}[ht]
\caption{Comparison of Experimental Results on Noisy Data using Noise Tolerant PHEC and PHEC in Centralized Settings} % title of Table
\centering % used for centering table
\begin{tabular}{c c c c} % centered columns (4 columns)
\hline %inserts double horizontal lines
Dataset & $\rho$ & Accuracy of PHEC & Accuracy of NT-PHEC\\ [0.5ex] % inserts table
%heading
\hline % inserts single horizontal line
NSL-KDD & 10 & 82.18 & \textbf{83.22}\\
NSL-KDD & 20 & 76.36 & \textbf{81.02}\\
DS2OS & 10 & 85.96 & \textbf{99.96}\\
DS2OS & 20 & 79.25 & \textbf{95.13}\\
Gas Pipeline & 10 & 90.86 & \textbf{97.48}\\
Gas Pipeline & 20 & 84.24 & \textbf{92.74}\\
Water Tank & 10 & 87.05 & \textbf{97.83}\\
Water Tank & 20 & 86.95 & \textbf{93.74}\\
[1ex] % [1ex] adds vertical space
\hline %inserts single line
\end{tabular}
\end{table}

\noindent
The following observations can be made on noisy data:
\begin{enumerate}
    \item It can be observed in Table-\textrm{XXIII} that for different degrees of noise, the accuracy obtained by Noise Tolerant PHEC is better than that obtained from PHEC for each of the four datasets. On noisy data, mean difference of the accuracy obtained by NT-PHEC and PHEC in centralized settings is 8.53\% (approx).
    \item As evident from Table-\textrm{XXIV}, performance metrics obtained by NT-PHEC is better than that obtained by PHEC in federated settings. On noisy data, the mean difference of accuracy (across four datasets, for various degrees of noise) between NT-PHEC and PHEC in federated settings is 2.6735\% (approx).
    
\end{enumerate}

\begin{table}[ht]
\caption{Comparison of Experimental Results on Noisy Data using Noise Tolerant PHEC and PHEC in Federated Settings} % title of Table
\centering % used for centering table
\begin{tabular}{c c c c} % centered columns (4 columns)
\hline %inserts double horizontal    lines
Dataset & $\rho$ & Accuracy of PHEC & Accuracy of NT-PHEC\\ [0.5ex] % inserts table
%heading
\hline % inserts single horizontal line
NSL-KDD & 10 &  72.87 & \textbf{76.48}\\
NSL-KDD & 20 & 68.28 & \textbf{74.22}\\
DS2OS & 10 & 89.9 & \textbf{91.66}\\
DS2OS & 20 & 88.46 & \textbf{90.48}\\
Gas Pipeline & 10 & 91.36 & \textbf{94.26}\\
Gas Pipeline & 20 & 89.98 & \textbf{93.988}\\
Water Tank & 10 & 89.82 & \textbf{90.47}\\
Water Tank & 20 & 89.72 & \textbf{90.22}\\
[1ex] % [1ex] adds vertical space
\hline %inserts single line
\end{tabular}
\end{table}
\noindent
On clean data, PHEC performs better than NT-PHEC, in federated and centralized settings. The mean difference of accuracy between PHEC and NT-PHEC in centralized and federated settings are 2.1\% \& and 4\% respectively (approx.).
\begin{table}[ht]
\caption{Comparison of Experimental Results on Clean Data using Noise Tolerant PHEC and PHEC in Federated and Centralized Settings} % title of Table
\centering % used for centering table
\begin{tabular}{c c c c} % centered columns (4 columns)
\hline %inserts double horizontal lines
Dataset & Settings & Accuracy of PHEC & Accuracy of NT-PHEC\\ [0.5ex] % inserts table
%heading
\hline % inserts single horizontal line
NSL-KDD & Centralized & \textbf{92.05} & 83.77\\
DS2OS & Centralized & \textbf{99.98} & 99.97\\
Gas Pipeline & Centralized & \textbf{98.38} & 98.33\\
Water Tank & Centralized & \textbf{98.71} & 98.64\\
NSL-KDD & Federated &                \textbf{88.42} & 83.65\\
DS2OS & Federated & \textbf{98.27} & 94.224\\
Gas Pipeline & Federated & \textbf{98.71} & 94.75\\
Water Tank & Federated & \textbf{93.74} & 90.48\\
[1ex] % [1ex] adds vertical space
\hline %inserts single line
\end{tabular}
\end{table}

\begin{enumerate}
    \item In centralized settings, PHEC uses Random Forest instead of weighted Logistic Regression. Random Forest performs better on clean data due to ensembling of results. For NT-PHEC, by imposing the constraint $K \geq 5$ on the KNN classifier, the feasible region for $K$ becomes a subset of the feasible region for $K$ of KNN in PHEC. So, the optimal $K$ of KNN in PHEC is atleast as good as that of KNN in NT-PHEC. Due to these reasons, PHEC performs better than NT-PHEC on clean data.  
    \item For the NT-PHEC in federated settings, the MLP is having one input layer and one output layer with single neuron. There exists no hidden layer in this case. The MLP in PHEC has the luxury of using multiple hidden layers with any number of neurons along-with any non-linear activation functions. Due to the large extent of flexibility of MLP, it captures the pattern of the data better than a simplified MLP (of NT-PHEC) and hence PHEC outperforms NT-PHEC on clean data.
\end{enumerate}

\section{Conclusion} 
\label{sec:Conclusion}
\noindent
We propose a probabilistic hybrid ensemble classifier (PHEC) in IoT security in centralized settings and then adapt it to federated settings. PHEC maximizes TPR while keeping the FPR within permissible limit. Experimental results demonstrate that it performs better than state-of-the-art Intrusion Detection Systems. Even though the results obtained in centralized setup are better than that of federated settings, performance metrics obtained in federated settings is quite close to centralized settings. It is advisable to use PHEC in the centralized settings if there exists no privacy issue and entire data processing in a single system is feasible. Otherwise, it is advisable to adapt the proposed model PHEC in federated settings. The classifiers used in PHEC use the training data labels to learn the pattern from the data. If labels of the training data get noisy, performance of PHEC drops. To overcome this, we use weighted convex surrogate loss functions to propose Noise-Tolerant PHEC in federated \& centralized settings. Experimental results on noisy data demonstrate that Noise-Tolerant PHEC works well in the presence of SLN Noise. 

An interesting direction of future work may be to come up with Intrusion Detection Systems such that it can be trained even in the absence of anomaly samples in the training data, the one class models may be useful there. 
\bibliographystyle{IEEEtran}
\bibliography{ref} 

% Generated by IEEEtran.bst, version: 1.14 (2015/08/26)
\begin{thebibliography}{10}
\providecommand{\url}[1]{#1}
\csname url@samestyle\endcsname
\providecommand{\newblock}{\relax}
\providecommand{\bibinfo}[2]{#2}
\providecommand{\BIBentrySTDinterwordspacing}{\spaceskip=0pt\relax}
\providecommand{\BIBentryALTinterwordstretchfactor}{4}
\providecommand{\BIBentryALTinterwordspacing}{\spaceskip=\fontdimen2\font plus
\BIBentryALTinterwordstretchfactor\fontdimen3\font minus
  \fontdimen4\font\relax}
\providecommand{\BIBforeignlanguage}[2]{{%
\expandafter\ifx\csname l@#1\endcsname\relax
\typeout{** WARNING: IEEEtran.bst: No hyphenation pattern has been}%
\typeout{** loaded for the language `#1'. Using the pattern for}%
\typeout{** the default language instead.}%
\else
\language=\csname l@#1\endcsname
\fi
#2}}
\providecommand{\BIBdecl}{\relax}
\BIBdecl

\bibitem{TDTC_2019}
H.~Haddad~Pajouh, R.~Javadian, R.~Khayamiand, A.~Dehghantanha, and R.~Choo,
  ``{A two layer dimension reduction and two tier classification model for
  anomaly based intrusion detection in IoT backbone networks},'' \emph{IEEE
  Transactions on Emerging Topics in Computing}, vol.~7, no.~2, 2016.

\bibitem{TwoTier_2015}
H.~H. Pajouh, G.~Dastghaibyfard, and S.~Hashemi, ``{Two-tier network anomaly
  detection model: a machine learning approach},'' \emph{J. Intell. Inf.
  Syst.}, pp. 1--14, 2015.

\bibitem{Rhman_20}
S.A.Rahman, H.~Tout, C.~Talhi, and A.Mourad, ``{Internet of Things Intrusion
  Detection: Centralized, On-Device, or Federated Learning?}'' \emph{IEEE
  Network}, 2020.

\bibitem{Evil_Tradeoff}
R.~V. Craiu and L.~Sun, ``{Choosing the lesser evil: trade-off between false
  discovery rate and non-discovery rate},'' \emph{Statistica Sinica.}, vol.~18,
  no.~3, pp. 861--879, 2008.

\bibitem{Tradeoff_2}
A.~W. Cappelen, C.~Cappelen, and B.~Tungodden, ``{Second-Best Fairness Under
  Limited Information: The Trade-Off between False Positives and False
  Negatives},'' \emph{Discussion Paper Series in Economics, Norwegian School of
  Economics}, 2018.

\bibitem{Natarajan_0}
N.~Natarajan, I.~S. Dhillon, P.~Ravikumar, and A.~Tewari, ``{Learning with
  Noisy Labels},'' \emph{NIPS}, 2013.

\bibitem{NAIVE_BAYES_2012}
L.~Koc, T.~A. Mazzuchi, and S.~Sarkani, ``{A network intrusion detection system
  based on a Hidden Naïve Bayes multiclass classifier},'' \emph{Expert Syst.
  Appl.}, vol.~39, no.~18, p. 13492–13500, 2012.

\bibitem{detailedKDD2019}
M.~Tavallaee, E.~Bagheri, W.~Lu, and A.~A. Ghorbani, ``{A detailed analysis of
  the KDD CUP 99 data set},'' \emph{Proceedings of the IEEE Symposium on
  Computational Intelligence for Security and Defence Applications}, 2009.

\bibitem{Zhang_RF_2008}
J.~Zhang, M.~Zulkernine, and A.~Haque, ``{Random-Forests based network
  intrusion detection systems},'' \emph{IEEE Trans. Syst. Man Cybern. Part C
  Appl. Rev.}, vol.~38, no.~5, p. 649–659, 2008.

\bibitem{Casas_unsupervised_2012}
P.~Casas, J.~Mazel, and P.~Owezarski, ``{Unsupervised network intrusion
  detection systems: Detecting the unknown without knowledge},'' \emph{Comput.
  Commun.}, vol.~35, no.~7, pp. 772--783, 2012.

\bibitem{SVM_BIRCH_HORNG}
S.~J. Horng \emph{et~al.}, ``{A novel intrusion detection system based on
  hierarchical clustering and support vector machines},'' \emph{Expert Syst.
  Appl.}, vol.~38, no.~1, pp. 306--313, 2011.

\bibitem{Ambusaidi_2016}
M.~A. Ambusaidi, X.~He, P.~Nanda, and Z.~Tan, ``{Building an Intrusion
  Detection System Using a FilterBased Feature Selection Algorithm},''
  \emph{IEEE Trans. Comput.}, vol.~65, no.~10, pp. 2986--2998, 2016.

\bibitem{Xuren_2006}
W.~Xuren, H.~Famei, and X.~Rongsheng, ``{Modeling Intrusion Detection System by
  Discovering Association Rule in Rough Set Theory Framework},''
  \emph{International Conference on Computational Intelligence for Modelling,
  Control and Automation}, p.~24, 2006.

\bibitem{Grouping_NSL_KDD}
S.~Revathi and D.~A. Malathi, ``{A Detailed Analysis on NSL-KDD Dataset Using
  Various Machine Learning Techniques for Intrusion Detection},''
  \emph{International Journal of Engineering Research \& Technology}, vol.~2,
  no.~12, 2013.

\bibitem{osana_DDOS_2016}
O.~Osanaiye, K.~R. Choo, and M.~Dlodlo, ``{Distributed denial of service (DDoS)
  resilience in cloud: Review and conceptual cloud DDoS mitigation
  framework},'' \emph{J. Netw. Comput. Appl.}, vol.~67, no.~7, pp. 147--165,
  2016.

\bibitem{Fed_basics_concept_and_applications}
Q.~Yang, Y.~Liu, T.~Chen, and Y.~Tong, ``{Federated Machine Learning: Concept
  and Applications},'' \emph{ACM Trans. Intell. Syst. Technol.}, 2019.

\bibitem{Challenges_Fed_Learning}
P.~Kairouz \emph{et~al.}, ``{Advances and Open Problems in Federated
  Learning},'' \emph{arXiv preprint}, 2021.

\bibitem{NSL_DATA}
``{NSL-KDD Dataset},'' \url{https://github.com/defcom17/NSL\_KDD}.

\bibitem{Federated_Tuhin}
\BIBentryALTinterwordspacing
T.~Sharma. (2019) Anomaly detection in smart buildings using federated
  learning. [Online]. Available:
  \url{https://www.tuhinsharma.com/talks/oreillyailondon2019/}
\BIBentrySTDinterwordspacing

\bibitem{Natarajan}
N.~Natarajan, I.~S. Dhillon, P.~Ravikumar, and A.~Tewari, ``{Cost-Sensitive
  Learning with Noisy Labels},'' \emph{Journal of Machine Learning Research},
  2018.

\bibitem{Aritra_Ghosh_2015}
A.~Ghosh, N.~Manwani, and P.~Sastry, ``{Making risk minimization tolerant to
  label noise},'' \emph{Neurocomputing}, 2015.

\bibitem{Naresh_2013}
N.~Manwani and P.~S. Sastry, ``{Noise Tolerance Under Risk Minimization},''
  \emph{IEEE Transactions on Cybernetics}, 2013.

\bibitem{CCN_2019}
H.~Reeve and A.~Kaban, ``{Classification with unknown class-conditional label
  noise on non-compact feature spaces},'' \emph{Conference on Learning Theory
  (COLT)}, 2019.

\bibitem{NOISE_knn}
S.~Ougiaroglou and G.~Evangelidis, ``{Dealing with noisy data in the context of
  k-NN Classification},'' \emph{BCI '15: Proceedings of the 7th Balkan
  Conference on Informatics}, 2015.

\bibitem{NSL_data_resource}
{Canadian Institute of Cybersecurity}, ``{NSL-KDD Dataset},''
  \url{https://www.unb.ca/cic/datasets/nsl.html}, 2009.

\bibitem{Dataset_2}
\BIBentryALTinterwordspacing
M.O.Pahl and F.~Aubet. (2018) Ds2os traffic traces. [Online]. Available:
  \url{https://www.kaggle.com/francoisxa/ds2ostraffictraces}
\BIBentrySTDinterwordspacing

\bibitem{DATA3_14}
W.~T.~Morris, ``{Industrial Control System Network Traffic Data sets to
  Facilitate Intrusion Detection System Research},'' \emph{IFIP Advances in
  Information and Communication Technology}, 2014.

\bibitem{D'Orazio_2016}
C.~D'Orazio, K.-K.~R. Choo, and L.~T. Yang, ``{Data Exfiltration from Internet
  of Things Devices: iOS Devices as Case Studies},'' \emph{IEEE Internet of
  Things Journal}, 2016.

\bibitem{HFRMLR_2016}
``{A Novel Anomaly Detection System Based on HFR-MLR Method},'' \emph{Mobile,
  Ubiquitous, and Intelligent Computing}, vol. 274, pp. 279--286, 2016.

\bibitem{NEURO_FUZZY_Toosi}
A.~N. Toosi and M.~Kahani, ``{A new approach to intrusion detection based on an
  evolutionary soft computing model using neuro-fuzzy classifiers},''
  \emph{Comput. Commun.}, vol.~30, no.~10, pp. 2201--2212, 2007.

\bibitem{bangladesh_19}
M.~M. M.Hasan, M.M.Islam, ``{Attack and anomaly detection in IoT sensors in IoT
  sites using machine learning approaches},'' \emph{IFIP Advances in
  Information and Communication Technology}, 2019.

\end{thebibliography}
\end{document}